\documentclass[%
 aip,
 amsmath,amssymb,
 reprint,%
]{revtex4-1}
\usepackage{comment}
\usepackage{textcomp}
\usepackage{graphicx}
\usepackage{dcolumn}
\usepackage{bm}

\usepackage[utf8]{inputenc}
\usepackage[T1]{fontenc}
\usepackage{mathptmx}

\begin{document}

\preprint{AIP/123-QED}

\title[]{Superconducting microwire detectors with single-photon sensitivity in the near-infrared}

\author{J. Chiles}
 \email{jeffrey.chiles@nist.gov}
\affiliation{ 
NIST, 325 Broadway, Boulder, CO 80305, USA
}%

\author{S. M. Buckley}
\affiliation{ 
NIST, 325 Broadway, Boulder, CO 80305, USA
}%

\author{A. Lita}
\affiliation{ 
NIST, 325 Broadway, Boulder, CO 80305, USA
}%

\author{V. B. Verma}
\affiliation{ 
NIST, 325 Broadway, Boulder, CO 80305, USA
}%

\author{J. Allmaras}
\affiliation{ 
JPL, 4800 Oak Grove Dr, Pasadena, CA 91109, USA
}%

\author{B. Korzh}
\affiliation{ 
JPL, 4800 Oak Grove Dr, Pasadena, CA 91109, USA
}%

\author{M. D. Shaw}
\affiliation{ 
JPL, 4800 Oak Grove Dr, Pasadena, CA 91109, USA
}%

\author{J. M. Shainline}
\affiliation{ 
NIST, 325 Broadway, Boulder, CO 80305, USA
}%

\author{R. P. Mirin}
\affiliation{ 
NIST, 325 Broadway, Boulder, CO 80305, USA
}%

\author{S. W. Nam}
\affiliation{ 
NIST, 325 Broadway, Boulder, CO 80305, USA
}%

\date{\today}

\begin{abstract}
We report on the fabrication and characterization of single-photon-sensitive WSi superconducting detectors with wire widths from 1 \textmu m to 3 \textmu m.  The devices achieve saturated internal detection efficiency at 1.55 \textmu m wavelength and exhibit maximum count rates in excess of 10\textsuperscript{5} s\textsuperscript{-1}. We also investigate the material properties of the silicon-rich WSi films used for these devices. We find that many devices with active lengths of several hundred microns exhibit critical currents in excess of 50\% of the depairing current.  A meandered detector with 2.0 \textmu m wire width is demonstrated over a surface area of 362 $\times$ 362 \textmu m\textsuperscript{2}, showcasing the material and device quality achieved.
\end{abstract}

\maketitle

Superconducting nanowire single-photon detectors (SNSPDs) \cite{Goltsman2001,hoiv2019} have become invaluable in a broad range of applications thanks to their superior detection efficiency \cite{Marsili2013,EsmaeilZadeh2017,Smirnov2018,Reddy2019}, low dark count rates \cite{hochberg2019detecting}, and ease of integration with integrated photonic circuits \cite{Buckley2017,mcdonald2019}.  Thanks to the small superconducting energy gap of many films used in these devices, they can exhibit a large optical detection bandwidth reaching out to the mid-IR \cite{Marsili2012,Marsili2013a,Chen2018,Verma2019}.  Typically, wire widths on the order of 100 nm are necessary to achieve saturated internal detection efficiency (IDE) in the infrared.  However, recent experiments by Korneeva \textit{et al.} have shown that microbridge-type MoSi superconducting detectors can exhibit single-photon sensitivity up to $\lambda$ = 1 \textmu m for wire widths of 1 \textmu m \cite{Korneeva2018a}.  It is thought that a prerequisite for this behavior is a high material homogeneity and minimal geometric constrictions in the device, permitting bias currents close to the intrinsic depairing current of the unpatterned film \cite{Vodolazov2017,Korneeva2018}, at which point a hotspot generated in a detection event can lead to a normal region across the entire wire, independent of the wire width. Simultaneously, the superconducting gap of the material also affects the number of Cooper pairs broken in an absorption event, leading to more efficient detection for wide wires with a lower-gap material.  Exploiting these two behaviors would enhance the competitiveness of SNSPDs fabricated with photolithography \cite{Shainline2017a}, allowing high-yield production in a parallel process.  Furthermore, wide wires on the micron scale would naturally carry much more current and enhance the signal generated from a detection event (although trade-offs for modifying the superconducting gap must also be considered), potentially reducing the cost and complexity of the electronics that are required per detector channel.  A larger signal, in general, also leads to lower jitter for nanowire detectors.

While the results in Ref. \onlinecite{Korneeva2018a} are highly encouraging and have stimulated new interest in exploring the physics of detection events in SNSPDs, there are significant caveats.  One is that a detection efficiency plateau is not observed for wavelengths longer than 1 \textmu m, and the other is that the devices demonstrated to date are still only short bridges, suggesting that the intrinsic material inhomogeneity is still a limiting factor in realizing reliable `microwire' devices.  The study of microwire devices would benefit from parallel efforts in alternative amorphous superconducting materials such as W\textsubscript{x}Si\textsubscript{1-x}, which exhibits high material homogeneity allowing excellent yield and consistent performance across many devices. Recently, we demonstrated kilopixel detector arrays with yield > 99\% \cite{wollman2019kilopixel}. In this manuscript, we fabricate microwire and nanowire devices in several WSi films of different stoichiometry and thickness and compare their performance.  We also study the fundamental material properties of these films.  We demonstrate saturated IDE at $\lambda$ = 1.55 \textmu m in out-and-back microwire detectors with wire widths up to 3 \textmu m and wire lengths of 100 \textmu m, as well as in meandered microwire devices (2 \textmu m wire width) over a detector area of 362 $\times$ 362 \textmu m\textsuperscript{2} (10,000 squares of wire).  For contrast, the typical size of meandered detectors is on the order of 15 $\times$ 15 \textmu m\textsuperscript{2}.
\begin{figure}[b]
	\centering
	\includegraphics[width=1\linewidth]{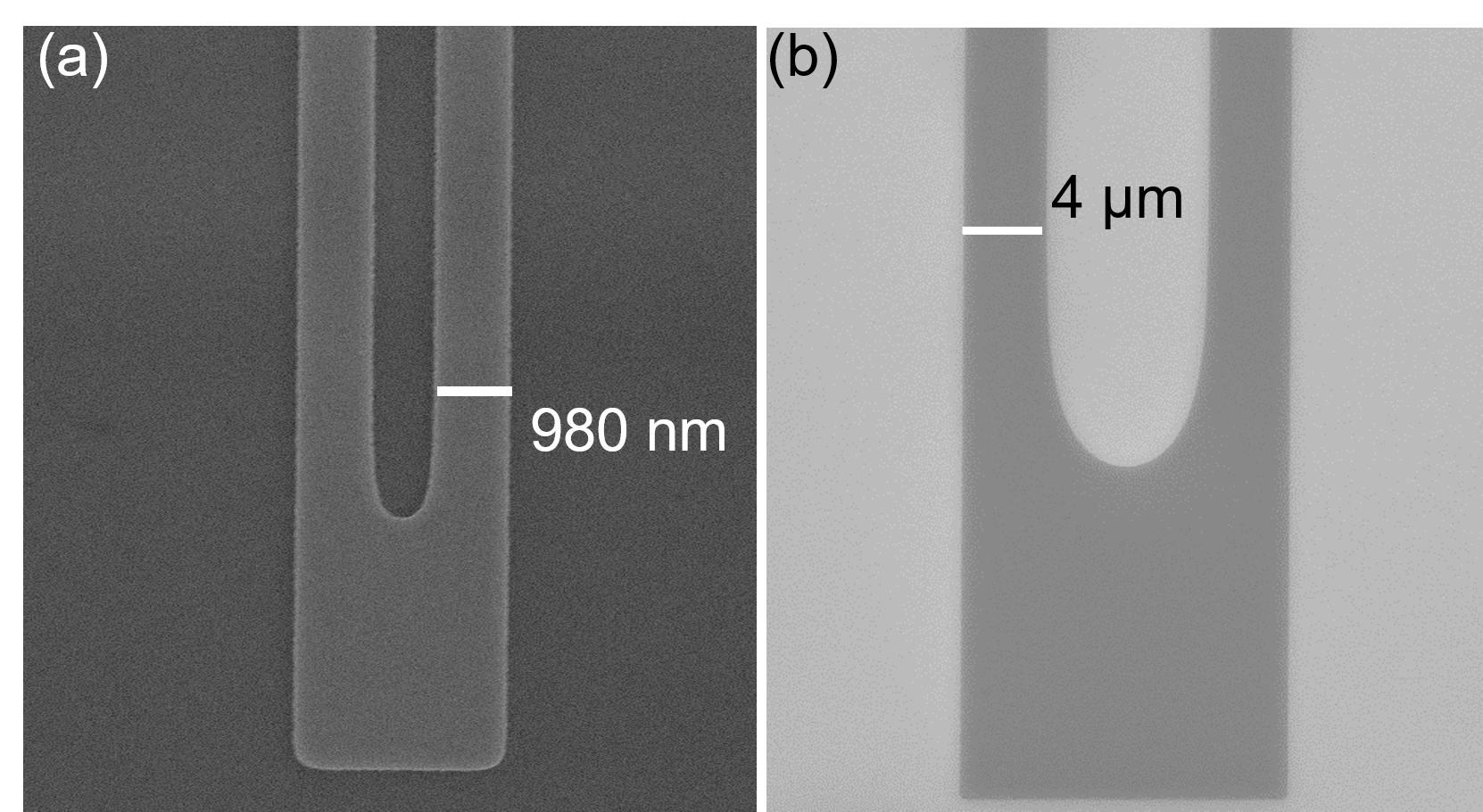}
	\caption{(a) SEM of an out-and-back style microwire device fabricated with photolithography.  (b) A wider microwire device fabricated with electron-beam lithography.}
	\label{fig:geometry}
\end{figure}

\begin{figure*}[tb]
	\centering
	\includegraphics[width=1.0\linewidth]{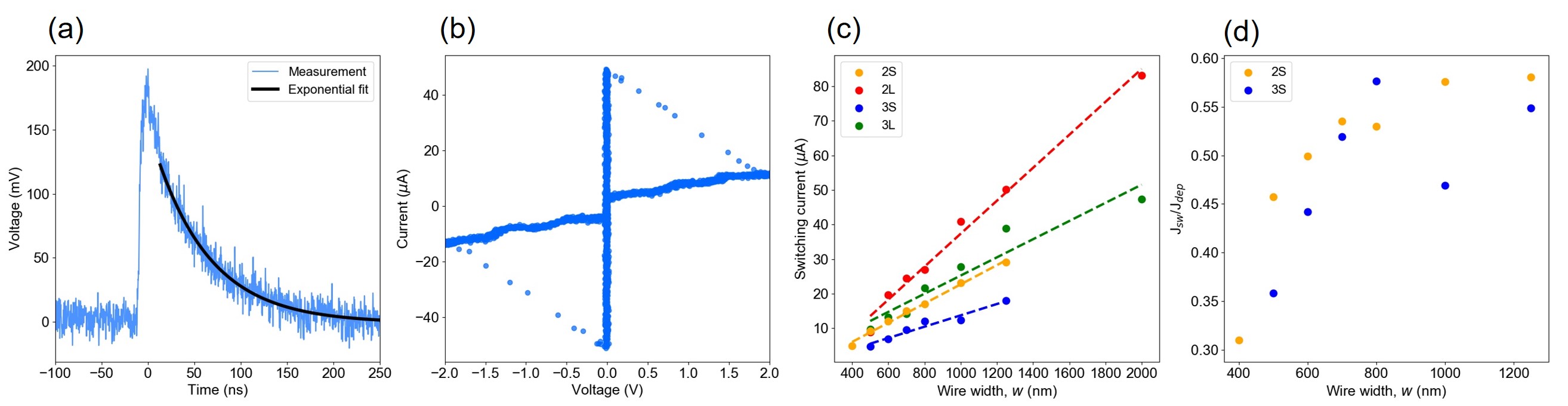}
	\caption{(a) Time-voltage trace of an infrared photon detection event in a typical device, showing the fitted exponential used to determine the decay time and kinetic inductance. (b) Example I-V trace for determining switching currents. (c) Trend of the switching current vs. wire width for the four film types. (d) Trend of the switching current to depairing current ratio vs. wire width. }
	\label{fig:iv}
\end{figure*}

We considered several variations on film properties in the process of fabricating and testing devices used in this work.  In the process of optimizing a film for a given application, the superconducting energy gap is often tailored to optimize or trade-off the superconducting transition temperature (T\textsubscript{c}), the critical current density, and the sensitivity to long-wavelength photons \cite{Verma2019,Zichi2019}.  In the case of WSi, the superconducting energy gap can be reduced by increasing the fraction of silicon in the film, and reducing the thickness.  In this work, we compare four types of WSi films in a 2 $\times$ 2 matrix: either 260 W or 320 W of sputtering power applied during sputtering of the silicon, and a deposited film thickness of either 2.1 nm or 2.8 nm.  The film properties are compared in Table~\ref{tab:material_param}.

\begin{figure*}[t]
	\centering
	\includegraphics[width=0.95\linewidth]{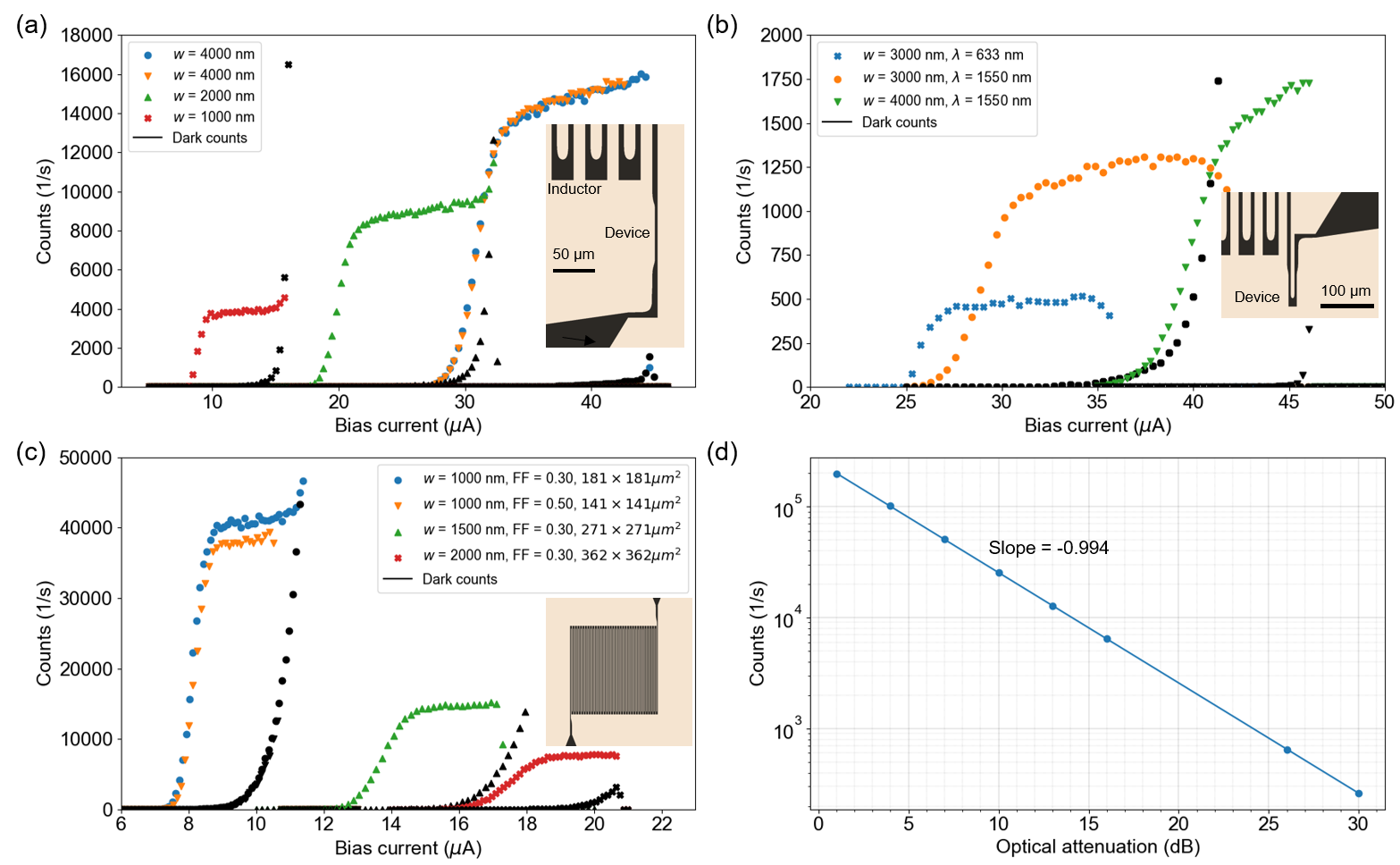}
	\caption{Measured results from the final e-beam-patterned 3S-type film. (a-c): Photon and dark count rates for (a) straight wires 65 \textmu m long, (b) out-and-back wires of 100 \textmu m total length, and (c) meandered detectors of various widths and areas.  The reduced absolute count rates for the right-most traces are because data were collected with different conditions and sample locations.  Note that dark counts have been subtracted from the photon count rate traces.  Insets of (a-c): layout for each type of device geometry. (d) Linearity measurement showing good performance over 3 orders of magnitude in count rates for a 2 \textmu m wide out-and-back device.}
	\label{fig:lastfig}
\end{figure*}

The devices were fabricated at the 76-mm-diameter wafer scale by depositing WSi onto a silicon wafer with 160 nm thermal oxide, followed by lithographic patterning and plasma etching with dilute SF\textsubscript{6} in Ar. The wafer was then diced into 1 cm-square chips which were wire-bonded to sample mounts and cooled to 750-800 mK.  Most of the devices were patterned with conventional 365 nm photolithography, except for a final fabrication run conducted with electron-beam (e-beam) lithography for the 3S-type film. An example microwire out-and-back device after fabrication is shown in Fig. \ref{fig:geometry}(a).  The various device geometries used in this manuscript are also shown in the insets of Fig.~\ref{fig:lastfig}(a-c): straight wire, out-and-back, and meander.  On-chip meandered inductors (with wires significantly wider than the width of the device under test) are used to prevent latching.

\begin{table}[b]
\caption{\label{tab:material_param}Comparison of material parameters for the four types of films used in this experiment.}
\begin{ruledtabular}
\begin{tabular}{ccccc}
Sample name&\mbox{2S}&\mbox{2L}&\mbox{3S}&\mbox{3L}\\
\hline
Thickness (nm)&\mbox{2.1}&\mbox{2.8}&\mbox{2.1}&\mbox{2.8}\\
Si mol. frac.&\mbox{0.36 $\pm 0.07$}&\mbox{0.36 $\pm 0.07$}&\mbox{0.48 $\pm 0.1$}&\mbox{0.48 $\pm 0.1$}\\
T\textsubscript{c} (K)&\mbox{3.2}&\mbox{3.5}&\mbox{2.9}&\mbox{3.3}\\
$\Delta$\textsubscript{0} (meV)&\mbox{0.49}&\mbox{0.53}&\mbox{0.44}&\mbox{0.50}\\
L\textsubscript{k},exp. (pH/$\square$)&\mbox{280}&\mbox{190}&\mbox{420}&\mbox{230}\\
J\textsubscript{dep} (MA/cm\textsuperscript{2})&\mbox{1.5}&\mbox{-}&\mbox{1.2}&\mbox{-}\\        
\end{tabular}
\end{ruledtabular}
\end{table}

For each sample type, we conducted measurements on the as-deposited films to determine various intrinsic properties, summarized in Table \ref{tab:material_param}. The critical temperature is defined here as the temperature at which the sheet resistance of the film reaches 50\% of the sheet resistance at 20 K. The film stoichiometry was determined by secondary ion mass spectroscopy.  The kinetic inductance was determined by fitting a single-exponential decay to the voltage response of the detector, as shown in Fig.~\ref{fig:iv}(a).  
\begin{figure*}[t]
	\centering
	\includegraphics[width=0.95\linewidth]{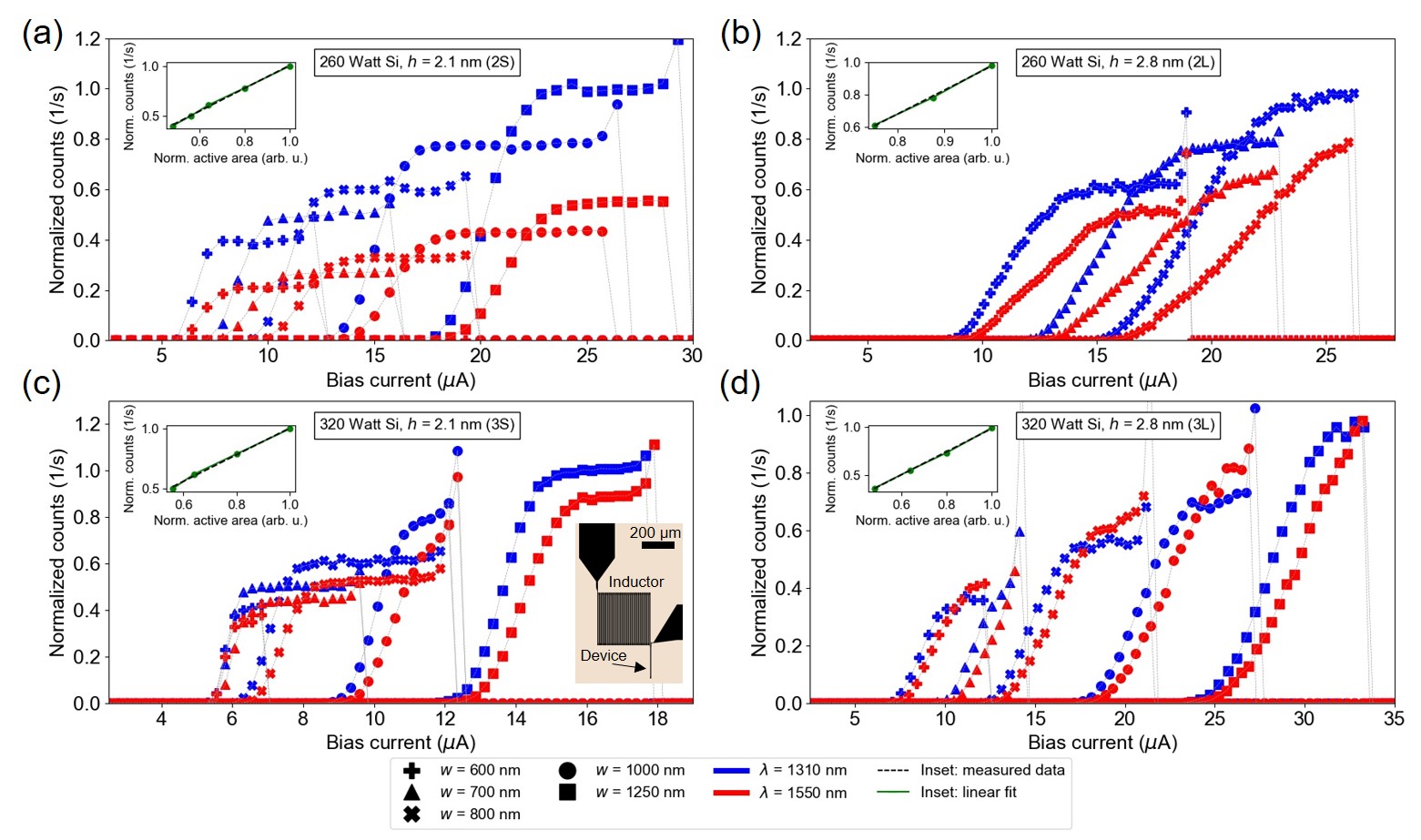}
	\caption{Photon count rates vs. bias current for the (a) 2S, (b) 2L, (c) 3S, and (d) 3L samples for different wire widths indicated in the legend below.  Note that dark counts have not been subtracted in this data set but are visible as rising tails near the critical current in some cases. Small bumps within the plateau regions are due to amplifier saturation artifacts. Upper-left insets: Normalized count rate vs. active area for each detector, showing the expected linear trend and validating that counts are originating from the straight sections and not constricted areas.  The lower-right inset of (c) shows the device geometry for these tests.}
	\label{fig:cvb_4wafers}
\end{figure*}
Before testing the photon-detection performance, we characterized the critical current performance of the devices and different films.  An example I-V curve is plotted in Fig.~\ref{fig:iv}(b).  We performed I-V sweeps for wire widths from 400 nm to 1250 nm and for all four sample types; the switching current results are collected in Fig.~\ref{fig:iv}(c). While they mostly follow linear fits to the nanowire width as one would expect, it is clear that some deviation from the ideal values is present when the data are instead normalized to the depairing current density, as in Fig.~\ref{fig:iv}(d) for samples 2S and 3S.  In both cases, the ratio of switching current density $J_{sw}$ to depairing current density $J_{dep}$ is lower for wire widths of < 700 nm, but mostly saturates for larger wires, most likely due to small constrictions from photolithographic patterning. Regardless, the wider devices exhibit $J_{sw}/J_{dep}$ > 0.55, despite the significant total wire length of 400 \textmu m and the presence of a hairpin turn in the device structure.

Next, we characterized the photon detection performance of out-and-back detectors for wire widths up to 1250 nm and a fixed total wire length of 400 \textmu m, for all four film types (Fig.~\ref{fig:cvb_4wafers}(a-d)).  The geometry is shown in the inset of Fig.~\ref{fig:cvb_4wafers}(a).  Flood illumination was used to characterize the device responses to photons at 1310 nm and 1550 nm. A typical count rate of $10^{3}-10^{4}$ s\textsuperscript{-1} was recorded during these measurements. Rates varied somewhat between samples due to non-uniform illumination, and are normalized in this particular measurement to allow their straightforward comparison.  For films 2S, 2L and 3S, good plateaus in the IDE are seen for $\lambda$ = 1310 nm for most wire widths in the range that we tested, and films 2S and 3S exhibited plateaus at 1550 nm as well.  Comparing the thinner to the thicker films, it suggests the reduced thickness is quite beneficial to photon detection performance in wide wires here, which is reasonable to expect given the change in superconducting energy gap.




So far in this work, we have only examined microwire devices with widths up to 1.25 \textmu m. We also fabricated much wider microwire devices in both straight wire and out-and-back configurations (see insets of Fig.~\ref{fig:lastfig} for visual reference). Devices had widths from 1 \textmu m to 4 \textmu m. A separate fabrication run was conducted using the 3S-type film and e-beam patterning. The measurement results of all devices from this run are collected in Fig.~\ref{fig:lastfig}.  Note that a slightly lower T\textsubscript{c} of 2.5 K was seen in this sample due to temporary issues with the deposition system, but wide devices made on the previous 3S-type film also showed similar saturation curve results.  First, we consider the photon detection performance of straight microwires in Fig.~\ref{fig:lastfig}(a) of lengths of 65 \textmu m.  Under 1550 nm illumination, the 3 \textmu m and 4 \textmu m-wide devices achieve a plateau-like region with a small linear slope that is likely due to counts in the tapers connecting the device to the bond-pad and inductor. For the out-and-back type devices (Fig.~\ref{fig:lastfig}(b)) with a longer total length of about 100 \textmu m, we observed a clear plateau in the IDE under 1550 nm illumination for a wire width of 3 \textmu m.  For the 4 \textmu m wide out-and-back device, the count rate is seen to just begin to flatten although a clear plateau is not observed.  The final device geometry tested was meandered detectors. Fig.~\ref{fig:lastfig}(c) shows the photon counting performance for several such detectors, with widths of 1, 1.5 and 2 \textmu m and fill fractions of 30\% or 50\%. All these devices show a clear plateau in their response to 1550 nm light despite wire lengths of 10,000 squares.  In the case of the 2 \textmu m wide-wire meander occupying 362 $\times$ 362 \textmu m\textsuperscript{2}, its surface area is 0.13 mm\textsuperscript{2}.  Finally, to confirm that the measurements of microwires was performed in a linear regime, we tested a 2 \textmu m wide out-and-back device of 100 \textmu m total length with variable optical attenuation (Fig.~\ref{fig:lastfig}(d)).  The count rate is highly linear for count rates of $2\times10^{2}$ s\textsuperscript{-1} to  $2\times10^{5}$ s\textsuperscript{-1}.

In conclusion, we report the fabrication and characterization of single-photon-sensitive microwire detectors with widths ranging from 1 \textmu m to 3 \textmu m and lengths ranging from 65 \textmu m to 20,000 \textmu m.  The material properties of different WSi film thicknesses and stoichiometries are compared to motivate and explain the performance we achieve in the experiments.  A meandered detector with a wire width of 2 \textmu m and an area of 362 $\times$ 362 \textmu m\textsuperscript{2} is shown to have saturated IDE at $\lambda$ = 1550~nm, demonstrating that the fabrication process and the material homogeneity are both sufficiently high-quality for state-of-the-art detector fabrication in this platform.  These results have implications for the high-volume production of large-area single-photon detectors with potential applications in dark matter detection \cite{hochberg2019detecting}, single-photon optical metrology \cite{fox1991trap}, and imaging arrays \cite{wollman2019kilopixel}.

\begin{acknowledgments}
We thank Karl K. Berggren and Ilya Charaev at MIT for helpful discussions.  This is a contribution of NIST, an agency of the U.S. government, not subject to copyright.
\end{acknowledgments}

\appendix

\nocite{*}
\bibliography{aipsamp}
\end{document}